\documentclass{article}
\usepackage{amsfonts}
\usepackage{eurosym}
\usepackage{amssymb}
\usepackage{amsmath}
\usepackage{cite}

\begin{document}

\title{Symmetries and conservation laws for the generalized $n$-dimensional
Ermakov system}
\author{Andronikos Paliathanasis\thanks{%
Email: anpaliat@phys.uoa.gr} \thanks{%
Coresponding Author} \\
{\ \textit{Institute of Systems Science, Durban University of Technology }}\\
{\ \textit{PO Box 1334, Durban 4000, Republic of South Africa}} \\
{\textit{Instituto de Ciencias F\'{\i}sicas y Matem\'{a}ticas,}}\\
{\ \textit{Universidad Austral de Chile, Valdivia, Chile}}, \and Genly Leon
\\
{\ \textit{Departamento de Matem\'{a}ticas, Universidad Cat\'{o}lica del
Norte, Avda. }}\\
{\ \textit{A}}\textit{ngamos 0610, Casilla 1280 Antofagasta, Chile } \\
{\ \textit{Institute of Systems Science, Durban University of Technology }}\\
{\ \textit{PO Box 1334, Durban 4000, Republic of South Africa}} \and P.G.L.
Leach \\
{\ \textit{Institute of Systems Science, Durban University of Technology }}\\
{\ \textit{PO Box 1334, Durban 4000, Republic of South Africa}} }
\maketitle

\begin{abstract}
We revise recent results on the classification of the generalized
three-dimensional Hamiltonian Ermakov system. We show that a statement
published recently is incorrect, while the solution for the classification
problem was incomplete. We present the correct classification for the
three-dimensional system by using results which related the background space
with the dynamics. Finally, we extend our results for the generalized $n$%
-dimensional Hamiltonian Ermakov system. \newline
\newline

\end{abstract}

\section{Introduction}

\label{sec1}

At the end of the 19th century, Ermakov \cite{Ermakov} solved the problem
for the derivation of first integral for the time-dependent linear equation
\begin{equation}
\ddot{x}+\omega ^{2}(t)x=0,  \label{AEP.00.1}
\end{equation}%
where an overdot means total derivative with respect to the time-variable, $%
t $, that is, $\dot{x}=\frac{dx}{dt}$.

By introducing the new variable $\rho \left( t\right) $ which satisfies the
second-order differential equation
\begin{equation}
\ddot{\rho}+\omega ^{2}(t)\rho =\rho ^{-3}  \label{AEP.00.2}
\end{equation}%
the time-dependent term in (\ref{AEP.00.1}) can be eliminated and the
following function $J\left( \rho ,\dot{\rho},x,\dot{x}\right) $ can be
constructed%
\begin{equation}
J\left( \rho ,\dot{\rho},x,\dot{x}\right) =\frac{1}{2}\left[ (\rho \dot{x}-%
\dot{\rho}x)^{2}+(x/\rho )^{2}\right] ,  \label{AEP.00.3}
\end{equation}%
which is a conservation law, i.e. $\dot{J}=0$. The second-order ordinary
differential equation (\ref{AEP.00.2}) was solved by Pinney in \cite{Pinney}%
. The analytic solution is
\begin{equation}
\rho ^{2}\left( t\right) =c_{1}\left( \rho _{1}\left( t\right) \right)
^{2}+c_{2}\left( \rho _{2}\left( t\right) \right) ^{2}+2c_{3}\rho _{1}\left(
t\right) \rho _{2}\left( t\right) ,  \label{ep.01}
\end{equation}%
where $\rho _{1}\left( t\right) $ and $\rho _{2}\left( t\right) $ are
solutions of equation (\ref{AEP.00.1}) and $c_{1}$,~$c_{2}$ and $c_{3}$ are
constants, of which only two are independent. Hence, equation (\ref{AEP.00.2}%
) is known as the Ermakov-Pinney equation. Furthermore, in \cite{Lewis},
Lewis investigated the integral properties for the time-dependent harmonic
oscillator, of equation (\ref{AEP.00.1}) in classical and quantum systems.
Lewis rediscovered the the Ermakov invariant (\ref{AEP.00.3}) and function $%
J\left( \rho ,\dot{\rho},x,\dot{x}\right) $ nowadays is called Ermakov-Lewis
invariant. The oscillator appears in many physical problems and other areas
of Applied Mathematics. Consequently, the Ermakov-Lewis invariant and the
Ermakov-Pinney equation have appeared in many studies, from quantum
mechanics \cite{Lewis,q1,q2}, to gravitational theory \cite{q4,q5} and
others \cite{q7,q8,q9,q10}.\ Recently, in \cite{pand1} the physical
interpretation of the Ermakov-Lewis invariant was discussed.

The application of the Lie theory on the Ermakov-Pinney equation provides
that equation (\ref{AEP.00.2}) admits three Lie point symmetries which form
the $SL(2,R)$ algebra \cite{le1}. The representation of the Lie algebra
depends upon the term $\omega ^{2}\left( t\right) .$ Thus, there exists a
point transformation in which the term $\omega ^{2}\left( t\right) $ can be
eliminated and equation (\ref{AEP.00.2}) to be reduced in the simple form%
\begin{equation}
\ddot{\rho}=\rho ^{-3}.  \label{ep.02}
\end{equation}

There are many extensions and generalization of the Ermakov-Pinney equation
in the literature. The two-dimensional system,%
\begin{eqnarray}
\ddot{x}+\omega ^{2}\left( t\right) x &=&\frac{f\left( \frac{y}{x}\right) }{%
x^{3}}  \label{ep.03} \\
\ddot{y}+\omega ^{2}\left( t\right) y &=&\frac{g\left( \frac{y}{x}\right) }{%
y^{3}},  \label{ep.04}
\end{eqnarray}%
is known as an Ermakov system \cite{ray}. System (\ref{ep.03}), (\ref{ep.04}%
) admits the conservation law \cite{le2}
\begin{equation}
I\left( x,\dot{x},y,\dot{x}\right) =\frac{1}{2}\left( x\dot{y}-\dot{x}%
y\right) ^{2}+\int^{y/x}\left( uf\left( u\right) -u^{-3}g\left( u\right)
\right) du
\end{equation}%
which is a generalization of the Ermakov-Lewis invariant. We remark that
there exist a point transformation such that the system (\ref{ep.03}), (\ref%
{ep.04}) can be written in an equivalent form without the $\omega ^{2}\left(
t\right) $ term. \ Moreover, the two-dimensional Ermakov system admits three
point symmetries which form the the $SL\left( 2,R\right) $ algebra. In the
special case for which $f\left( \frac{y}{x}\right) =V\left( x,y\right) _{,x}$
and $g\left( \frac{y}{x}\right) =V\left( x,y\right) _{y}$, the dynamical
system follows from a variational principle and the Lie symmetries are also
Noether symmetries.

By using the latter property for the Ermakov system, that is, the admitted
Lie symmetries for a given dynamical system to form the $SL\left( 2,R\right)
$ algebra, there have been various generalizations of the Ermakov system in
higher-dimensions \cite{le1}, in curved geometries \cite{aner1} or many
others see for instance \cite{ga1,ga2,ga3,ga4} and references therein. The
three-dimensional Ermakov system introduced in \cite{le1}, for which it was
found that in the case that the Ermakov system is Hamiltonian the
Ermakov-Lewis invariant is a function of the variational conservation laws
which are determined by the applications of Noether's second theorem for the
elements of the $SL\left( 2,R\right) $ algebra, see also the discussion in
\cite{hass2}. However, such a property was found also in the case of curved
geometries. Moreover, for the Hamiltonian Ermakov system, the Ermakov-Lewis
invariant can be constructed by Noether's second theorem for contact
symmetries. Indeed, there was found to be an exact relation for the
existence of the Killing tensor which provides the\ Ermakov-Lewis invariant
and of the main generator vector for the background space of the $SL\left(
2,R\right) $ Lie algebra \cite{aner1}.\ The $SL\left( 2,R\right) $ Lie
algebra plays an important role in the study of the integrability of various
Hamiltonian systems, we refer the reader in some recent studies \cite%
{ll1,ll2,ll3,ll4,ll5,ll6,ll7}

A complete classification of variational symmetries for the Hamiltonian
three-dimensional Ermakov \cite{le1} system performed in \cite{naz}.\ The
authors determined the functional form for the unknown potential functions
for the generalized Ermakov system to admit additional variational
symmetries which produce conservation laws except from the elements of the $%
SL\left( 2,R\right) $ algebra. The authors determined eight different
potential functions of which they claim that four of these functions are new
in the literature and new exact solutions were determined. However, this
claim is false. As we shall see with a simple construction method the
potential functions the authors determined in \cite{naz} are well known in
the literature. Nevertheless, inspired by the problem which the authors of
\cite{naz} study, we extend our analysis and we find a general result for
the generalized Hamiltonian Ermakov system in $n-$dimensional Euclidian
geometry. The plan of the paper is as follow.

In Section \ref{sec2} we present basic properties and definitions of the
symmetry analysis. In Section \ref{sec3} we discuss the two-dimensional
Ermakov system and the Lewis invariant. The analysis of the
three-dimensional generalized Ermakov system is presented in Section \ref%
{sec4}. Finally, in Section \ref{sec5} we draw our conclusions.

\section{Preliminaries}

\label{sec2}

In this Section we present the basic properties and definitions for the
theory of symmetries of differential equations. We assume a system of
second-order ordinary differential equations of the form%
\begin{equation}
\ddot{x}^{i}=\Omega ^{i}\left( t,x^{j},\dot{x}^{j}\right) ~,  \label{Lie.0}
\end{equation}%
in which $x^{i}$ are the dependent variables and $t$ is the independent
variable. Furthermore, consider the one-parameter point transformation $\Phi
\left( t,x;\varepsilon \right) $ with infinitesimal generator $\xi \left(
t,x^{j}\right) \partial _{t}+\eta ^{i}\left( t,x^{j}\right) \partial _{i}$\
defined in the augmented space $\{t,x^{i}\}$. Thus, we say that the vector
field $X$ is a Lie point symmetry for the system (\ref{Lie.0}) if the set of
differential equations remains invariant under the action of the
one-parameter point transformation $\Phi \left( t,x;\varepsilon \right) $,
equivalently when the following condition is true \cite{StephaniB}
\begin{equation}
X^{\left[ 2\right] }\left( \ddot{x}^{i}-\Omega ^{i}\left( t,x^{j},\dot{x}%
^{j}\right) \right) =0~,  \label{Lie.1}
\end{equation}%
where $X^{\left[ 2\right] }$\ is the second prolongation of $X$ defined by
the formula $X^{\left[ 2\right] }=X+\left( \dot{\eta}^{i}-\dot{x}^{i}\dot{\xi%
}\right) \partial _{\dot{x}^{i}}+\left( \ddot{\eta}^{i}-\dot{x}^{i}\ddot{\xi}%
-2\ddot{x}^{i}\dot{\xi}\right) \partial _{\ddot{x}^{i}}$. An equivalent way
to express the symmetry condition (\ref{Lie.1}) is \cite{StephaniB}
\begin{equation}
\left[ X^{\left[ 1\right] },A\right] =\lambda \left( x^{k}\right) A~,
\label{Lie.3}
\end{equation}%
in which now $X^{\left[ 1\right] }$ is the first prolongation of $X$ and $A$
is the Hamiltonian vector field $A=\partial _{t}+\dot{x}\partial _{x}+\Omega
^{i}\left( t,x^{j},\dot{x}^{j}\right) \partial _{\dot{x}^{i}}.$

If the system of differential equations results from a variational
principle, that is, there exists a Lagrangian function $L=L\left( t,x^{j},%
\dot{x}^{j}\right) ,$\ then the infinitesimal generator $X$\ for the
one-parameter point transformation is characterized as variational symmetry
or as Noether symmetry for the given Lagrangian function, if there exists a
function $G$ such that the condition of Noether's first theorem is satisfied
\cite{noe}
\begin{equation}
X^{\left[ 1\right] }L+L\frac{d\xi }{dt}=\frac{dG}{dt}  \label{Lie.5}
\end{equation}%
Function $G$ is boundary term introduced to allow for the infinitesimal
changes in the value of the Action Integral produced by the infinitesimal
change in the boundary of the domain caused by the infinitesimal
transformation of the variables in the Action Integral.

The importancy of the determination of Noether symmetries for a given system
is that there exist a unique relation between Noether symmetries and
conservation laws. Indeed, if $X$ is a Noether symmetry vector with
corresponding boundary term $G$, then according to Noether's second theorem
the following function%
\begin{equation}
I\left( t,x^{j},\dot{x}^{j}\right) =\xi \left( \dot{x}^{i}\frac{\partial L}{%
\partial \dot{x}^{i}}-L\right) -\frac{\partial L}{\partial \dot{x}^{i}}\eta
^{i}+G  \label{Lie.6}
\end{equation}%
is a conservation law for the dynamical system, that is $\dot{I}\left(
t,x^{j},\dot{x}^{j}\right) =0$. For a recent review on Noether's theorems we
refer the reader to \cite{nrev}.

\section{The Ermakov system}

\label{sec3}

In the following we consider the Ermakov system and without loss of
generality we can omit the linear term with the time-dependent component $%
\omega ^{2}\left( t\right) $.

As we mentioned before the Ermakov-Pinney equation (\ref{ep.02}) admits
three Lie point symmetries, the vector fields
\begin{equation}
X^{1}=\partial _{t}~,~X^{2}=2t\partial _{t}+\rho \partial _{\rho }\text{~}%
,~X^{3}=t^{2}\partial _{t}+t\rho \partial _{\rho }~.  \label{ep.10}
\end{equation}%
Equation (\ref{ep.02}) admits the Lagrangian function
\begin{equation}
L\left( \rho ,\dot{\rho}\right) =\frac{1}{2}\left( \dot{\rho}^{2}-\rho
^{-2}\right) .  \label{eq.11}
\end{equation}%
Thus, by replacing the vector fields $\left\{ X^{1},X^{2},X^{3}\right\} $ in
Noether's condition we find that the three vector fields are Noether
symmetries with corresponding boundary terms $G\left( X^{1}\right) =G_{0}$, $%
G\left( X^{2}\right) =0$ and $G\left( X^{3}\right) =\frac{1}{2}\rho ^{2}$
respectively.

Hence, from expression (\ref{Lie.6}) we define the conservation laws%
\begin{eqnarray}
I_{1}\left( X^{1}\right) &=&\frac{1}{2}\left( \dot{\rho}^{2}+\rho
^{-2}\right) \equiv \mathcal{H},  \label{eq.12} \\
I_{2}\left( X^{2}\right) &=&2t\mathcal{H}-\rho \dot{\rho}~,~  \label{eq.14}
\\
I_{3}\left( X^{3}\right) &=&t^{2}\mathcal{H}-t\rho \dot{\rho}+\frac{1}{2}%
\rho ^{2}~,  \label{eq.15}
\end{eqnarray}%
where $\mathcal{H}$ is the Hamiltonian function.

Consider now the two-dimensional Hamiltonian Ermakov system%
\begin{equation}
\ddot{x}=\frac{\left( V\left( \frac{y}{x}\right) \right) _{x}}{x^{3}},~\ddot{%
y}=\frac{\left( V\left( \frac{y}{x}\right) \right) _{y}}{y^{3}},
\label{eq.16A}
\end{equation}%
or in polar coordinates $\left( x,y\right) =\left( r\cos \theta ,r\sin
\theta \right) $,
\begin{eqnarray}
\ddot{\rho}-\rho \dot{\theta}^{2} &=&\frac{V\left( \theta \right) }{\rho ^{3}%
}~,  \label{eq.16} \\
\ddot{\theta}+2\rho \dot{\theta} &=&\frac{V\left( \theta \right) _{,\theta }%
}{2\rho ^{4}}.  \label{eq.17}
\end{eqnarray}%
with Lagrangian function%
\begin{equation}
L\left( \rho ,\dot{\rho},\theta ,\dot{\theta}\right) =\frac{1}{2}\left( \dot{%
\rho}^{2}+\rho ^{2}\dot{\theta}^{2}-\frac{V\left( \theta \right) }{\rho ^{2}}%
\right) .  \label{eq.18}
\end{equation}

For arbitrary potential function $V\left( \theta \right) $ the dynamical
system (\ref{eq.16}), (\ref{eq.17}) admits as Lie point symmetries the
vector fields $\left\{ X^{1},X^{2},X^{3}\right\} $ which are also Noether
symmetries for the Lagrangian function (\ref{eq.18}). The corresponding
conservation laws are the $I_{1}\left( X^{1}\right) $, $I_{2}\left(
X^{2}\right) $ and $I_{3}\left( X^{3}\right) $ where now
\begin{equation}
\mathcal{H}=\frac{1}{2}\left( \dot{\rho}^{2}+\rho ^{2}\dot{\theta}^{2}+\frac{%
V\left( \theta \right) }{\rho ^{2}}\right) .
\end{equation}

By using these three functions we can define the following expression \cite%
{le1}%
\begin{equation}
J\left( \rho ,\dot{\rho}\right) =4I_{3}\left( X^{1}\right) \mathcal{H}%
-\left( I_{2}\left( X^{2}\right) \right) ^{2}  \label{eq.19}
\end{equation}%
which is nothing else than the Ermakov-Lewis invariant, that is
\begin{equation}
J\left( \rho ,\dot{\rho}\right) =\frac{1}{2}\left( \dot{\rho}^{2}+\rho
^{-2}\right) .  \label{eq.20}
\end{equation}

However, for specific forms of $V\left( \theta \right) $ additional Noether
symmetries can exist. Indeed when $V\left( \theta \right) =V_{0}$, equation (%
\ref{eq.17}) reads $\left( \mathcal{L}\right) ^{\cdot }=0$, where $\mathcal{L%
}=\rho ^{2}\dot{\theta}$ is the angular momentum. The additional Noether
symmetry in this case is the vector field $X^{4}=\partial _{\phi }$. The
functional forms of $V\left( \theta \right) $ for which additional
variational symmetries exist are presented in \cite{an2d}. We omit the
presentation of the symmetry vector and we present the potential functions.

The potentials are $V_{A}\left( \theta \right) =V_{0}$, $V_{B}\left( \theta
\right) =V_{0}\left( \cos \theta \right) ^{-2}$, $V_{C}\left( \theta \right)
=V_{0}\left( \sin \theta \right) ^{-2}$,~$V_{D}\left( \theta \right)
=V_{1}\left( \cos \theta \right) ^{-2}+V_{2}\left( \sin \theta \right) ^{-2}$
and $V_{E}\left( \theta \right) =V_{0}\left( V_{1}\sin \theta -\cos \theta
\right) ^{-2}$. We remark that potentials $V_{B}\left( \theta \right) $ and $%
V_{C}\left( \theta \right) $ are the same potential function, because $\sin
\theta =\cos \left( \theta -\frac{\pi }{2}\right) $. Thus there are only
four potentials. As we discussed, potential $V_{A}\left( \theta \right) $
provides the conservation law of angular momentum.

For potential $V_{B}\left( \theta \right) $, if we go into Cartesian
coordinates the Lagrangian function for the Ermakov system (\ref{eq.16A}) is%
\begin{equation}
L_{B}\left( x,\dot{x},y,\dot{y}\right) =\frac{1}{2}\left( \dot{x}^{2}+\dot{y}%
^{2}-\frac{V_{0}}{x^{2}}\right) ~,
\end{equation}%
which provides the second-order differential equations $\ddot{x}=\frac{V_{0}%
}{x^{3}}$~,~$\ddot{y}=0.$ The time-independent conservation laws are $%
J\left( x,\dot{x}\right) =\frac{1}{2}\left( \dot{x}^{2}+V_{0}x^{-2}\right) $
and $I^{2}=\dot{y}^{2}$. The potential $V_{C}\left( \theta \right) $ gives
the same conservation laws with $\left( x,y\right) \rightarrow \left(
y,x\right) $.

For potential $V_{D}\left( \theta \right) $ in Cartesian coordinates the
Lagrangian function becomes%
\begin{equation}
L_{D}\left( x,\dot{x},y,\dot{y}\right) =\frac{1}{2}\left( \dot{x}^{2}+\dot{y}%
^{2}-\frac{V_{1}}{x^{2}}-\frac{V_{2}}{y^{2}}\right) ~,
\end{equation}%
with equations of motion $\ddot{x}=\frac{V_{1}}{x^{3}}$ and $\ddot{y}=\frac{%
_{2}}{y^{3}}$, for which the conservation laws are $J_{1}\left( x,\dot{x}%
\right) =\frac{1}{2}\left( \dot{x}^{2}+V_{1}x^{-2}\right) $ and $J_{2}\left(
y,\dot{y}\right) =\frac{1}{2}\left( \dot{y}^{2}+V_{1}y^{-2}\right) $.

Finally, for the potential function $V_{E}\left( \theta \right) $ we have
the Lagrangian
\begin{equation}
L_{E}\left( x,\dot{x},y,\dot{y}\right) =\frac{1}{2}\left( \dot{x}^{2}+\dot{y}%
^{2}-\frac{V_{0}}{\left( \alpha y-x\right) ^{2}}\right).
\end{equation}%
By doing the second change of variables $x=X+\alpha y$ it follows that $%
L_{E}\left( X,\dot{X},y,\dot{y}\right) =\frac{1}{2}\left( \dot{X}%
^{2}+2\alpha \dot{X}\dot{Y}+\left( 1+\alpha ^{2}\right) \dot{y}^{2}-\frac{%
V_{0}}{X^{2}}\right) $ for which the equations of motion are $\ddot{X}=\frac{%
V_{0}\left( 1+\alpha ^{2}\right) }{X^{3}}$, $\ddot{y}=-\frac{\alpha V_{0}}{%
x^{3}}$. The time-independent conservation laws are $J\left( X,\dot{X}%
\right) =\frac{1}{2}\left( \dot{X}^{2}-V_{0}X^{-2}\right) $ and $I\left( X,%
\dot{X},\dot{Y}\right) =\left( 2\alpha \dot{X}\dot{Y}+\left( 1+\alpha
^{2}\right) \dot{y}^{2}\right) $.

We have seen that in all cases for which additional point symmetries exist
the Hamiltonian Ermakov system, when it is written in Cartesian coordinates,
the potential function $\frac{V\left( \theta \right) }{\rho ^{2}}$ becomes $%
V\left( x,y\right) =\frac{V_{1}}{x^{2}}+\frac{V_{2}}{y^{2}}$ and $V\left(
x,y\right) =\frac{V_{0}}{\left( \alpha y-x\right) ^{2}}$. In these cases the
additional point symmetries follow from the the gradient Killing symmetries
of the Euclidian space.

\section{The 3D generalized Ermakov system}

\label{sec4}

Consider now the three-dimensional Ermakov system with Lagrangian function
\begin{equation}
L\left( \rho ,\dot{\rho},\theta ,\dot{\theta},\phi ,\phi \right) =\frac{1}{2}%
\left( \dot{\rho}^{2}+\rho ^{2}\dot{\theta}+\rho ^{2}\sin ^{2}\theta ~\dot{%
\phi}^{2}-\frac{V\left( \theta ,\phi \right) }{\rho ^{2}}\right) ~
\label{eq.21}
\end{equation}%
and equations of motion
\begin{eqnarray}
\ddot{\rho}-\rho \dot{\phi}^{2}-\rho \sin ^{2}\theta ~\dot{\phi}^{2}-\frac{%
V\left( \theta ,\phi \right) }{R^{3}} &=&0,  \label{eq.22} \\
\ddot{\theta}+\frac{2}{\rho }\dot{\rho}\dot{\theta}-\sin \theta \cos \theta ~%
\dot{\phi}^{2}+\frac{V\left( \theta ,\phi \right) _{,\theta }}{2R^{4}} &=&0
\label{eq.23} \\
\ddot{\phi}+\frac{2}{\rho }\dot{\rho}\dot{\phi}+\cot \theta ~\dot{\theta}%
\dot{\phi}+\frac{V\left( \theta ,\phi \right) _{,\phi }}{2R^{4}\sin
^{2}\theta } &=&0.  \label{eq.24}
\end{eqnarray}

At this point we discuss the analysis presented in \cite{naz}. The authors
investigated the functional forms of $V\left( \theta ,\phi \right) $ for the
dynamical system to admit additional conservation laws based upon point
symmetries and contact symmetries. The authors determined the following
functional forms for the potential $V\left( \theta ,\phi \right) ,$%
\begin{equation}
V_{A}\left( \theta ,\phi \right) =V_{0}~,~V_{B}\left( \theta ,\phi \right)
=V_{0}\left( \cos \theta \right) ^{-2}\text{~,~}V_{C}\left( \theta ,\phi
\right) =V_{0}\left( \sin \theta \right) ^{-2}~,
\end{equation}%
\begin{equation}
V_{D}\left( \theta ,\phi \right) =V_{0}\left( \cos \theta \sin \phi \right)
^{-2}~,~V_{E}\left( \theta ,\phi \right) =V_{0}\left( \cos \theta \cos \phi
\right) ^{-2}~,
\end{equation}%
\begin{equation}
V_{F}\left( \theta ,\phi \right) =\frac{8}{\cos \left( 2\phi -2\theta
\right) +\cos \left( 2\phi +2\theta \right) +2\cos \left( 2\theta \right)
-2\cos \left( 2\phi \right) +6}~,
\end{equation}%
\begin{equation}
V_{G}\left( \theta ,\phi \right) =\frac{8}{2\cos \left( 2\phi \right) +2\cos
\left( 2\theta \right) -\cos \left( 2\phi -2\theta \right) -\cos \left(
2\phi +2\theta \right) +6}~.
\end{equation}

If we simplify potential functions $V_{F}$ and $V_{G}$ by using
trigonometric identities we find
\begin{equation}
V_{F}\left( \theta ,\phi \right) =\left( \cos ^{2}\phi \sin ^{2}\theta
-1\right) ^{-1},
\end{equation}%
\begin{equation}
V_{G}\left( \theta ,\phi \right) =\left( 1-\sin ^{2}\phi \sin ^{2}\theta
\right) ^{-1}.
\end{equation}%
As we discussed above the potentials are related, for instance $V_{B}$ and $%
V_{C}$ are the same potential function, as are also potentials $V_{D},~V_{E}$
and $V_{F},~V_{G}$. \ In order to understand that we should write the
dynamical systems in Cartesians coordinates. we assume the transformation%
\begin{equation}
\left( x,y,z\right) =\left( r\sin \theta ,r\cos \theta \sin \phi ,r\cos
\theta \cos \phi \right) \text{.}
\end{equation}%
Then
\begin{equation}
\rho ^{-2}V_{C}\left( \theta ,\phi \right) =x^{-2}~,~\rho ^{-2}V_{D}\left(
\theta ,\phi \right) =y^{-2}
\end{equation}%
and
\begin{equation}
\rho ^{-2}V_{F}\left( \theta ,\phi \right) =\left( x^{2}+z^{2}\right) ^{-1}~
\end{equation}%
while in a similar way the remaining potentials reduce to similar potentials
in Cartesians coordinates, by doing a change on the variables $\left\{
x,y,z\right\} $. These potentials are not new and have been found before and
widely studied in \cite{ts1}. In addition the classification scheme in \cite%
{naz} is incomplete, for which see below.

We want to omit any calculation and use results from the previous section as
also important properties of the symmetries for the Euclidian space $%
\mathbb{R}
^{3}$. There is a direct relation of the Noether symmetries with the
elements of the Homothetic algebra of the Euclidian space, for more details
we refer the reader in \cite{an2d}. The proper homothetic vector of the $%
\mathbb{R}
^{3}$ is the generator of the two vector fields of the $SL\left( 2,R\right) $
algebra. However, the $%
\mathbb{R}
^{3}$ space admits six isometries.\ Three rotations which form the $O\left(
3\right) $ Lie algebra and three translations which form the $T\left(
3\right) $ Lie algebra. From the nature of the $O\left( 3\right) $, if an
element of $O\left( 3\right) $ is a symmetry vector then we can change the
coordinates $\left( \theta ,\phi \right) \rightarrow \left( \bar{\theta},%
\bar{\phi}\right) ,$ the symmetry vector to be always the $\partial _{\bar{%
\phi}}$ which provides the conservation law for the angular momentum
\thinspace\ $\mathcal{L}=$ $\rho ^{2}\sin ^{2}\theta ~\dot{\phi}^{2}$. If we
assume that two elements of $O\left( 3\right) $ to be Noether symmetries,
then it follows that $V\left( \theta ,\phi \right) =V_{0}$.

Therefore, when $V\left( \theta ,\phi \right) =V_{A}^{1}\left( \theta
\right) $ or $V\left( \theta ,\phi \right) =V_{A}^{20}$, the additional
Noether symmetries follow from the elements of the $O\left( 3\right) $ Lie
algebra. On the other hand, if we assume that the symmetry is generated by
an element of the $T\left( 3\right) $ Lie algebra, it follows that in the
Cartesians coordinates the system is similar to that found above for the
two-dimensional system.

Indeed, we perform the symmetry classification%
\begin{equation}
V_{A}\left( x,y,z\right) =\frac{V_{1}}{x^{2}}+\frac{V_{2}}{y^{2}}+\frac{V_{3}%
}{z^{2}}~,
\end{equation}%
\begin{equation}
V_{B}\left( x,y,z\right) =\frac{V_{0}}{x^{2}+y^{2}}~+\frac{V_{1}}{z^{2}},
\end{equation}%
\begin{equation}
V_{E}=\frac{V_{0}}{\left( \alpha x-\beta y-\gamma z\right) ^{2}}\text{~,~~}%
V_{E}=\frac{V_{0}}{\left( \alpha x-\beta y\right) ^{2}}+\frac{V_{1}}{z^{2}}%
\,,
\end{equation}%
where $V\left( x,y,z\right) =\rho ^{-2}V\left( \theta ,\phi \right) $.

It is obvious that we have found new potentials which have not been
presented before in \cite{naz}. Furthermore, if we apply the Killing tensors
to determine new potential functions, from the algorithm described in \cite%
{ck1}; it is clear that there are no additional potential functions.

\section{Conclusions}

\label{sec5}

In this work we revised previous results on the classification problem of
the three-dimensional Hamiltonian Ermakov system by using point and contact
symmetries. We show that the classification problem solved in \cite{naz} is
not correct and incomplete. We were able to show that with a constructive
approach by using previous results which relate properties of the background
space with the dynamical system.

We show that there is a natural extension on the classification scheme from
the two-dimensional Ermakov system to the three-dimensional problem. Hence
we can use the analysis to elevate the solution in higher dimensions. We
summarize our analysis in the following proposition.

\textbf{Proposition:} \textit{Consider the }$n$\textit{-dimensional
Hamiltonian Ermakov system generated by the Lagrangian function}%
\begin{equation}
L\left( \rho ,\dot{\rho},\theta _{1},\dot{\theta}_{1},...,\theta _{n-1},\dot{%
\theta}_{n-1}\right) =\frac{1}{2}\left( \dot{\rho}^{2}+\rho ^{2}\left( \dot{%
\theta}_{1}^{2}+\sin ^{2}\theta _{1}\left( \dot{\theta}_{2}^{2}+\sin
^{2}\left( ...\right) \right) \right) -\frac{V\left( \theta _{1},\theta
_{2},...,\theta _{n-1}\right) }{\rho ^{2}}\right) ,
\end{equation}%
\textit{which admits the }$SL\left( 2,R\right) $\textit{\ as Noether
symmetries. The functional forms for the potential }$V\left( \theta
_{1},\theta _{2},...,\theta _{n-1}\right) $\textit{\ in which additional
conservation laws linear or quadratic in the momentum are:}%
\begin{equation}
V\left( \theta _{1},\theta _{2},...,\theta _{n-1}\right) =V_{0}
\end{equation}%
\begin{equation}
V\left( x_{1},x_{2},...,x_{n}\right) =\sum_{i}V_{i}\left( x^{i}\right) ^{-2}
\end{equation}%
\begin{equation}
V\left( x_{1},x_{2},...,x_{n}\right) =\sum_{i}\frac{V_{i}}{\sum_{\kappa
}x^{\kappa }x_{\kappa }}+\sum_{i}\frac{V_{i}}{\left( x^{\lambda }\right) ^{2}%
}~,~\kappa \neq \lambda \text{,}
\end{equation}%
\begin{equation}
V\left( x_{1},x_{2},...,x_{n}\right) =\sum_{i}\frac{V_{i}}{\sum_{\kappa
}\left( \alpha ^{\kappa }x_{\kappa }\right) ^{2}}+\sum_{i}\frac{\alpha
_{\lambda }V_{i}}{\left( x^{\lambda }\right) ^{2}}~,~\kappa \neq \lambda
\text{,~}
\end{equation}%
\textit{where }$\left( x_{1},x_{2},...,x_{n}\right) \,$\textit{\ are the
Cartesians coordinates for the }$n$\textit{-dimensional Euclidian space and }%
$\rho ^{-2}V\left( \theta _{1},\theta _{2},...,\theta _{n-1}\right) =V\left(
x_{1},x_{2},...,x_{n}\right) $.

In a future work we plan to extend this procedure to other attempts on the
generalization of the Ermakov system.

\subsection*{Acknowledgements}

GL was funded by Comisi\'{o}n Nacional de Investigaci\'{o}n Cient\'{\i}fica
y Tecnol\'{o}gica (CONICYT) through FONDECYT Iniciaci\'{o}n 11180126. GL
thanks the Department of Mathematics and the Vicerrector\'{\i}a de
Investigaci\'{o}n y Desarrollo Tecnol\'{o}gico at Universidad Cat\'{o}lica
del Norte for financial support. This work does not have any conflicts of
interest.

\end{document}